9
\documentclass[english,pra,aps,showpacs]{revtex4}
\usepackage{amssymb}

\makeatletter

\providecommand{\boldsymbol}[1]{\mbox{\boldmath $#1$}}

\usepackage{epsfig}
\usepackage{graphicx}
\usepackage{amssymb}
\usepackage{amsmath}

\usepackage{babel}
\makeatother
\begin{document}

\title{Simple Minimal Informationally Complete Measurements for Qudits}

\author{Stefan Weigert}

\affiliation{Department of Mathematics, University of York\\
 Heslington, UK-York YO10 5DD, United Kingdom\\
 \texttt{slow500@york.ac.uk}}

\date{August 2005}

\begin{abstract}
Simple minimal but informationally complete positive operator-valued
measures are constructed out of the expectation-value representation
for qudits. Upon suitable modification, the procedure transforms any
set of $d^{2}$ linearly independent hermitean operators into such
an observable. Minor changes in the construction lead to closed-form 
expressions for informationally complete positive measures in 
the spaces $\mathbb{C}^{d}$.
\end{abstract}


\maketitle

\newcommand{\ket}[1]{|#1\rangle}
 
\newcommand{\bra}[1]{\langle#1|}
 
\newcommand{\ketbra}[2]{|#1\rangle\langle#2|}

\section{Introduction\label{sec:Introduction}}

The very idea to implement information on quantum systems and to subsequently
process it \cite{nielsen+2000} requires to initially \emph{prepare}
a particular quantum state $\hat{\rho}_{in}$, to \emph{verify} the
preparation procedure and to \emph{identify} the final state $\hat{\rho}_{out}$
produced by the quantum dynamics. Since an unknown quantum state cannot
be determined unambiguously by a single measurement and no copies
of the state can be made, one needs to resort to repeated measurements
of identically prepared states. \emph{}It is thus desirable to design
quantum mechanical algorithms in such a way that only a small number
of \emph{known} final states can arise: it is easier then to extract
the desired information with high probability from a few runs as in
Shor's algorithm \cite{shor94}, for example. Both the verification
of a state and its identification are instances of \emph{state reconstruction}
or \emph{estimation} \cite{leonhardt97}.

A useful measure for the reliability of a measurement procedure to
determine an unknown quantum state is given by its \emph{fidelity}
$F$, the mean overlap of the reconstructed state with the exact state
\cite{uhlmann76}. Perfect fidelity usually requires an infinite supply
of the unknown state. This might sound unrealistic from an experimental
point of view. The theoretical possibility to achieve $F=1$ for an
arbitrary input state with a given set of measurements says that the
measurements are \emph{complete}. In the following, the focus will
be on sets of hermitean operators which allow one, in principle, to
perfectly reconstruct an unknown state described by a density operator
$\hat{\rho}$ in a $d$-dimensional Hilbert space $\mathcal{\mathbb{C}}^{d}$.
Many such bases for observables are known \cite{fano+59,brif+99},
and in some cases they have been combined into what is called \emph{minimal
informationally complete positive operator-valued measures} (cf. below)
which is important from a conceptual point of view. 

This contribution will strengthen the links between state reconstruction,
minimal complete sets of hermitean operators, and positive operator-valued
measures. Firstly, it argues that one can extract minimal informationally
complete measures from the \emph{expectation-value representation}
of quantum states in finite-dimensional Hilbert spaces \cite{weigert99/1}.
Secondly, this approach will be adapted to construct such measures
out of any set of $d^{2}$ linearly independent hermitean operators
on $\mathcal{\mathbb{C}}^{d}$.

\section{Positive Operator-valued Measures\label{sec: POVMs}}

\subsection{Properties and Examples\label{sub: general POVMs}}

Positive operator-valued measures \cite{kraus83} correspond to the
most general quantum mechanical observables. The following summary
collects their properties insofar as far as they are relevant here,
and some notation will be established. 

Consider a quantum system capable of residing in $d$ states $\ket{\psi_{n}},\: n=1\ldots d,$
which form an orthonormal basis of the $d$-dimensional Hilbert space
${\cal {H}}=\mathbb{C}^{d}$. A hermitean operator $\hat{E}=\hat{E}^{\dagger}$
is called \emph{positive semi-definite}, $\hat{E}\geq0,$ if there is no state which
produces a negative expectation value for $\hat{E}$, or equivalently,

\begin{equation}
\bra{\psi_{n}}\hat{E}\ket{\psi_{n}}\geq0\:,\quad n=1\ldots d\,.\label{eq: defpositivity}\end{equation}
 The density matrix $\hat{\rho}$ used to describe a mixed state of
a quantum system provides a well-known example of such an operator,
$\hat{\rho}\geq0$. A collection of positive operators $\hat{E}_{\alpha},\alpha\in A,$
with $A$ being a discrete or continuous set of labels, qualifies
as a \emph{positive operator-valued measure} in $\mathcal{H}$, or
POVM for short, if its elements sum up to the identity in $\mathcal{H}$,

\begin{equation}
\sum_{\alpha\in A}\hat{E}_{\alpha}=\mathbb{\hat{\mathbb{I}}}\:.\label{eq: povm unity}\end{equation}
When $\alpha$ is a continuous label, the symbol $\sum$ is understood
to denote an integration over $A$. Taking the expectation value of
this equation in any normalized state $\ket{\psi}$, one finds that
the discrete or continuous set of positive numbers $p_{\alpha}=\bra{\psi}\hat{E}_{\alpha}\ket{\psi},\alpha\in A,$
sum up to one. Thus, the numbers $p_{\alpha}$ have the properties
of a probability distribution which suggests to think of the operators
$\hat{E}_{\alpha},\alpha\in A,$ as an {}``operator-valued'' \emph{measure.} 

Here are four examples of POVMs. The first example consists of only
one element, the identity $\hat{\mathbb{I}}$ in $\mathcal{\mathbb{C}}^{d}$.
Next, the completeness relation of the states $\ket{\psi_{n}}$,\begin{equation}
\sum_{n=1}^{d}\ketbra{\psi_{n}}{\psi_{n}}=\hat{\mathbb{I}}\label{eq: psincompleteness}\end{equation}
 shows that the collection of the positive semi-definite, orthonormal
projectors $\hat{E}_{n}\equiv\ketbra{\psi_{n}}{\psi_{n}}$ form a
POVM with $d$ elements.

A POVM may contain any number of elements, \emph{}not restricted by
$d$, the dimension of the underlying Hilbert space. In such a situation,
the elements of the POVM cannot consist of orthonormal projections
since the space $\mathcal{H}$ accommodates at most $d$ orthogonal
states. Consider the example of a POVM for a qubit with Hilbert space
$\mathcal{\mathbb{C}}^{2}$, defined in terms of the states $\ket{\pm}$,
the eigenstates of the $z$-component of a spin $1/2$, equivalent
to the computational basis for the qubit. It consists of \emph{three}
operators,\begin{equation}
\hat{E}_{1}=\frac{\sqrt{2}}{1+\sqrt{2}}\ket{-}\bra{-}\:,\quad\hat{E}_{2}=\frac{\sqrt{2}}{1+\sqrt{2}}\left(\ket{-}-\ket{+}\right)\left(\bra{-}-\bra{+}\right)\:,\label{eq: threePOVM}\end{equation}
and $\hat{E}_{3}=\hat{\mathbb{I}}-\hat{E}_{1}-\hat{E}_{2}$, which
sum up to the identity. This POVM allows one to successfully differentiate
between non-orthogonal quantum states \cite{peres88}. Imagine that
you are being asked to find out whether you have been sent the state
$\ket{?}$ which could be either $\ket{+}$ or $(1/\sqrt{2})(\ket{-}+\ket{+})$.
Using the above POVM to perform a measurement on the unknown state
$\ket{?}$, you will find,in each run, an outcome associated with one
of the three operators given above. In the first case, associated
with $\hat{E}_{1}$, you know that the state provided cannot have
been $\ket{+}$ since $\bra{+}\hat{E}_{1}\ket{+}=0$; similarly, you
know that the unknown state must have been $\ket{+}$ if the measurement
outcome corresponds to $\hat{E}_{2}$ since only this state has a
non-zero component {}``along'' $\hat{E}_{2}$. If the third outcome
occurs, nothing can be said about $\ket{?}$. If you were to perform
a measurement with any two \emph{orthonormal} projections, you could draw no conclusions
about $\ket{?}$ from a single run. By invoking the
POVM defined in (\ref{eq: threePOVM}), however, it is possible to
extract the desired information from a \emph{single} run of an experiment
if either outcome 1 or 2 occur. 

The final example of a POVM has uncountably many elements: let\begin{equation}
\hat{E}_{\mathbf{n}}=\ketbra{\mathbf{n}}{\mathbf{n}}\:,\quad\ket{\mathbf{n}}\in\mathcal{S}\,,\label{eq:CS POVM member}\end{equation}
where $\ket{\mathbf{n}}$ is a coherent state of a spin $s\equiv(d-1)/2$,
the label $\mathbf{n}$ being a vector pointing from the origin to
the point $P_{\mathbf{n}}$ on the unit sphere $\mathcal{S}$ in $\mathbb{R}^{3}$.
The overcompleteness relation of the coherent states \cite{perelomov86}
implies that these operators are a indeed a POVM,\begin{equation}
\int_{\mathcal{S}}\:\hat{E}_{\mathbf{n}}\: d\mu(\mathbf{n})\equiv\frac{d}{4\pi}\:\int_{\mathcal{S}}\ketbra{\mathbf{n}}{\mathbf{n}}\: d\mathbf{n}=\hat{\mathbb{I}}\,.\label{eq: CS POVM}\end{equation}

\subsection{Minimal informationally complete POVMs\label{sub:MIC POVMs}}

If a POVM is to be \emph{informationally complete} (IC for short),
each set of probabilities $\rho_{\alpha},\alpha\in A,$ must identify
a unique density matrix $\hat{\rho}$ satisfying \begin{equation}
\rho_{\alpha}\equiv\mbox{Tr}\left[\hat{\rho}\hat{E}_{\alpha}\right]\,.\label{eq: rho POVM general}\end{equation}
 The first three examples of POVMs just described in the previous section
are not informationally complete while the coherent-state POVM defined in (\ref{eq: CS POVM})
is: only one \emph{}operator $\hat{A}$ is associated with a \emph{Q}-symbol
$A_{\mathbf{n}}=\bra{\mathbf{n}}\hat{A}\ket{\mathbf{n}}\equiv\mbox{Tr}[\hat{A}\hat{E}_{\mathbf{n}}]$
\cite{perelomov86}. 

\emph{Minimal} informationally complete POVMs, or MIC-POVMs, contain
the \emph{least} number of elements such that the probabilities in
(\ref{eq: rho POVM general}) determine a unique density matrix $\hat{\rho}$.
This requirement is equivalent to saying that the operators $\hat{E}_{\alpha}$
form a \emph{}(minimal) \emph{basis} of the vector space of hermitean
operators acting on the Hilbert space $\mathcal{\mathbb{C}}^{d}$.
Counting the number of real parameters necessary to parameterize all
such operators, conveniently represented as hermitean matrices of
size $(d\times d)$, one concludes that a MIC-POVMs will contain precisely
$d^{2}$ (linearly independent) elements. For convenience, the normalization
condition $\mbox{Tr}[\hat{\rho}]=1$ is often relaxed, so that density
matrices are indeed parameterized by a total of $d^{2}$ real numbers. 

Not every set of $d^{2}$ operators spanning the hermitean operators
on $\mathbb{C}^{d}$ is a POVM. To see this, let us look at the
example of a spin $1/2$, or qubit. Observables $\hat{A}$ have the
form \begin{equation}
\hat{A}=A_{0}\hat{\mathbb{I}}+\mathbf{A}\cdot\hat{\boldsymbol{\sigma}}\:,\label{eq: bloch}\end{equation}
 with a real number $A_{0}$ and a real three-component vector $\mathbf{A}$,
while $\hat{\boldsymbol{\sigma}}=(\hat{\sigma}_{x},\hat{\sigma}_{y},\hat{\sigma}_{z})$
denotes the spin operator. The four operators $(\hat{\mathbb{I}},\hat{\boldsymbol{\sigma}})$
do not constitute a POVM since the expectation values of each of the
operators $\hat{\sigma_{i}},i=x,y,z,$ range from $-1$ to $+1$.
However, all is not lost yet: the three indefinite operators turn
positive by adding the identity:\begin{equation}
0\leq\bra{\psi}\left(\hat{\mathbb{I}}+\hat{\sigma}_{i}\right)\ket{\psi}\leq2\:,\quad i=x,y,z\,.\label{eq:positivity}\end{equation}
 This observation makes it easy to construct examples of MIC-POVMs
for a qubit,\begin{equation}
\hat{E}_{n}=\frac{1}{4}\left(\hat{\mathbb{I}}+\mathbf{n}_{n}\cdot\hat{\boldsymbol{\sigma}}\right)\geq0\,,\quad n=1\ldots4,\quad\mbox{where}\quad\sum_{n=1}^{4}\mathbf{n}_{n}=0\,,\label{eq: tetrahedral POVM}\end{equation}
and the four unit vectors $\mathbf{n}_{n}$ must not lie in a plane
\cite{flammia+04},. 

The CFS-construction presented in \cite{caves+02} ascertains the
existence of MIC-POVMs for qudits living in $\mathcal{\mathbb{C}}^{d}$.
Consider any set of $d^{2}$ linearly independent positive definite
operators $\hat{F}_{\alpha}>0$, say, satisfying the relation \begin{equation}
\sum_{\alpha=1}^{d^{2}}\hat{F}_{\alpha}=\hat{G}>0\,.\label{eq: positive sum}\end{equation}
Being positive definite, the operator $\hat{G}$ has a unique, strictly
positive square root $\hat{G}^{\frac{1}{2}}$ the inverse of which,
$\hat{G}^{-\frac{1}{2}}$, exists as well. Thus, the transformation
$\hat{F}_{\alpha}\to\hat{E}_{\alpha}=\hat{G}^{-\frac{1}{2}}\hat{F}_{\alpha}\hat{G}^{-\frac{1}{2}}$
is invertible, preserves positivity, hermiticity and the rank of the
original operators. What is more, the new operators satisfy the relation
(\ref{eq: povm unity}) with $A=\{1\ldots d^{2}\}$, thus giving rise
to a MIC-POVM. As shown in \cite{caves+02} there is at least one
collection of $d^{2}$ linearly independent positive definite operators
in $\mathbb{C}^{d}$.

Particularly interesting examples of MIC-POVMs consist of projection
operators onto $d^{2}$ states $\ket{\varphi_{n}},n=1\ldots d^{2},$
such that their pairwise scalar products are of modulus $1/(d+1)$,
\begin{equation}
|\bra{\varphi_{n}}\varphi_{n^{\prime}}\rangle|^{2}=\frac{1}{d+1}\,,\quad n\neq n^{\prime}\,.\label{SIVPOVM}\end{equation}
It is known how to analytically construct such \emph{symmetric} MIC-POVMs,
or SIC-POVMs for short, in some Hilbert spaces of small dimensions as well as $d=19$ \cite{appleby04},
although numerical evidence up to $d=45$ \cite{renes+04} seems to suggest that they exist in any dimension (see \cite{problem23} for a survey).

\section{The Expectation-Value Representation of Quantum Mechanics\label{sec:EVR}}

\subsection{Definition of the Expectation-Value Representation\label{sub:Def EVR}}

If you randomly pick $d^{2}$ points $\mathbf{n}_{n},n=1\ldots d^{2},$
on the unit sphere, then the operators $\hat{Q}_{n}=\ketbra{\mathbf{n}_{n}}{\mathbf{n}_{n}}$,
projecting on the associated coherent states $\ket{\mathbf{n}_{n}}$
are, with probability one, linearly independent \cite{amiet+00}.
Consequently, they provide a basis for the hermitean operators on
$\mathcal{\mathbb{C}}^{d}$, \begin{equation}
\hat{A}=\frac{1}{d}\sum_{n=1}^{d^{2}}A^{n}\hat{Q}_{n}\:,\label{eq: expand operator}\end{equation}
with unique real coefficients $A^{n}$ (different from $\mbox{Tr}[\hat{A}\hat{Q}_{n}]$).
The trace of the product of two operators on $\mathcal{\mathbb{C}}^{d}$
defines a scalar product (one needs to invoke the adjoint of one of
the operators if non-hermitean operators are considered) which can
be used to introduce a second basis \emph{dual} to the projectors
$\hat{Q}_{n}$,

\begin{equation}
\frac{1}{d}\mbox{Tr}\left[\hat{Q}^{n}\hat{Q}_{n^{\prime}}\right]=\delta_{n^{\prime}}^{n}\,,\quad n,n^{\prime}=1\ldots d^{2}\,.\label{eq:duality}\end{equation}
The dual operators $\hat{Q}^{n},n=1\ldots d^{2},$ provide a basis
for observables just as the original ones do, \begin{equation}
\hat{A}=\frac{1}{d}\sum_{n=1}^{d^{2}}A_{n}\hat{Q}^{n}\:,\label{eq: EVR expansion}\end{equation}
with a second set of real expansion coefficients $A_{n}$. Using (\ref{eq:duality}),
one sees that the expansion coefficients in one basis are given by
the scalar product of the operator at hand with the corresponding
element of the dual basis, 

\begin{equation}
A^{n}=\mbox{Tr}\left[\hat{A}\hat{Q}^{n}\right]\:,\quad\mbox{and}\quad A_{n}=\mbox{Tr}\left[\hat{A}\hat{Q_{n}}\right]\ \,,\quad n=1\ldots d^{2}\,.\label{eq:dual coefficients}\end{equation}
It is interesting to point out that the coefficients $A_{n}$ and
$A^{n}$ can be thought of as discrete, non-redundant versions of
the \emph{Q}- and \emph{P}-symbols of the operator $\hat{A}$, respectively
\cite{amiet+00}. Knowing one set of coefficients, the other set is
determined uniquely by \begin{equation}
A_{n}=\frac{1}{d}\sum_{m=1}^{d^{2}}\mathfrak{G}_{nm}A^{m}\,,\label{eq: trf of coeffs}\end{equation}
where $\mathfrak{G}$ is the non-singular Gram matrix of the basis
$\hat{Q}_{n}$, with elements $\mathfrak{G}_{nm}=\mbox{Tr}[\hat{Q}_{n}\hat{Q}_{m}]$.
The coefficients $A_{n}$ have a simple physical meaning: recalling
that the $\hat{Q}_{n}$ are projections, one has \begin{equation}
A_{n}=\bra{\mathbf{n}_{n}}\hat{A}\ket{\mathbf{n}_{n}}\,,\label{eq:prob exp coeffs}\end{equation}
saying that any operator $\hat{A}$ is determined entirely by its
expectation values in $d^{2}$ appropriate coherent states. Consequently,
it is possible to parameterize the density matrix $\hat{\rho}$ of
a qudit (or a spin with $s=(d-1)/2$) by $d^{2}$ probabilities, $p_{n}=\bra{\mathbf{n}_{n}}\hat{\rho}\ket{\mathbf{n}_{n}}$.
These probabilities can be measured in $d^{2}$ independent experiments,
each corresponding to a different orientation of a standard Stern-Gerlach
apparatus \cite{weigert99/1}. When expressing a density matrix by
means of (\ref{eq: EVR expansion}) $\hat{\rho}$ is said to be given
in the \emph{expectation-value representation} (EVR for short).

\subsection{Obstacles}

Let us now explore whether the expectation-value representation gives
rise to MIC-POVMs. Being positive semi-definite, the $d^{2}$ operators
$\hat{Q}_{n}$ are promising candidates for a minimal informationally
POVM. But do they add up to the identity? There is an expansion of
the identity, \begin{equation}
\hat{\mathbb{I}}=\frac{1}{d}\sum_{n=1}^{d^{2}}\mathbb{I}^{n}\hat{Q}_{n}\:,\label{eq: identity in basis}\end{equation}
 and the rescaled projectors $(\mathbb{I}^{n}/d)\hat{Q}_{n}$ would
constitute a MIC-POVM if all coefficients $\mathbb{I}^{n}=\mbox{Tr}[\hat{Q}^{n}]$
were known to be positive. Unfortunately, the numbers $\mathbb{I}^{n}$
are \emph{not} guaranteed to be positive for all constellations of
directions $\{\mathbf{n}_{n},n=1\ldots d^{2}\}$, as follows from
the example presented in Section \ref{sub:Generic-MIC-POVMs}. 

In view of this result, it might be a good idea to expand the identity
in the \emph{dual} basis, \begin{equation}
\hat{\mathbb{I}}=\frac{1}{d}\sum_{n=1}^{d^{2}}\hat{Q}^{n}\:,\label{eq: dual evr expansion}\end{equation}
with automatically positive expansion coefficients $\mathbb{I}_{n}=\bra{\mathbf{n}_{n}}\,\hat{\mathbb{I}}\,\ket{\mathbf{n}_{n}}\equiv1$.
However, some of the dual operators $\hat{Q}^{n}$ will, in general,
not be positive semi-definite. To see this, consider the elements
of Gram matrix $\mathfrak{G}$ which are non-negative,\begin{equation}
\mathfrak{G}_{nn^{\prime}}=\mbox{Tr}\left[\hat{Q}_{n}\hat{Q}_{n^{\prime}}\right]=\left|\langle\mathbf{n}_{n}\ket{\mathbf{n}_{n^{\prime}}}\right|^{2}\geq0\,,\quad n,n^{\prime}=1.\ldots d^{2}\,;\label{eq: gram elements}\end{equation}
 the value zero is attained only if two vectors happen to point to
diametrically opposite points, $\mathbf{n}_{n}=-\mathbf{n}_{n^{\prime}}$,
implying that the Gram matrix has at most one zero in each row. It
follows that the inverse $\mathfrak{G}^{-1}$ of the Gram matrix must
have at least one \emph{negative} entry (actually, in each row): the
off-diagonal elements of the product $\mathfrak{G}^{-1}\mathfrak{G}$
could not vanish otherwise. Expressing the matrix elements of $\mathfrak{G}^{-1}$
by the scalar products of the elements of the dual basis, one is led
to conclude that \begin{equation}
\mathfrak{G}^{\nu\nu^{\prime}}=\mbox{Tr}\left[\hat{Q}^{\nu}\hat{Q}^{\nu^{\prime}}\right]<0\,,\label{eq: neg element of G-1}\end{equation}
for at least one pair of indices $\nu,\nu^{\prime}$, say. This relation
is incompatible with all operators $\hat{Q}^{n},n=1\ldots d^{2},$
being positive semi-definite: evaluate the trace in (\ref{eq: neg element of G-1})
in the eigenstates $\ket{Q_{r}^{\nu}},r=1\ldots d,$ of the operator
$\hat{Q}^{\nu}$ with eigenvalues $Q_{r}^{\nu}$. This implies \begin{equation}
\mbox{Tr}\left[\hat{Q}^{\nu}\hat{Q}^{\nu^{\prime}}\right]=\sum_{r=1}^{d}Q_{r}^{\nu}\bra{Q_{r}^{\nu}}\hat{Q}^{\nu^{\prime}}\ket{Q_{r}^{\nu}}<0\,,\label{eq: neg element of G-2}\end{equation}
which requires at least one negative term in the sum. Consequently,
either $\hat{Q}^{\nu}$ must have a negative eigenvalue or there is
a state such that the expectation value of $\hat{Q}^{\nu^{\prime}}$
is negative. Both alternatives show that not all dual operators $\hat{Q}^{\nu}$
can be positive semi-definite. Thus, one of the two operators in (\ref{eq: neg element of G-1})
is not be positive semi-definite, and the relation (\ref{eq: dual evr expansion})
does \emph{not} define a POVM.

\section{Constructing new MIC-POVMs}

The minimal informationally complete sets $\{\hat{Q}_{n},n=1\ldots d^{2}\}$
and $\{\hat{Q}^{n},n=1\ldots d^{2}\}$ will serve as starting points
to construct new MIC-POVMs.

\subsection{CFS-construction}

Let us apply the method by CFS to construct a POVM out of positive
multiples of the projection operators $\hat{Q}_{n}$. The sum \begin{equation}
\hat{S}=\sum_{n=1}^{d^{2}}\alpha_{n}\hat{Q}_{n}\:,\alpha_{n}>0\,,\label{eq: define S}\end{equation}
defines a hermitean, \emph{strictly} positive operator, $\hat{S}>0$,
as shown now. The expectation value of $\hat{S}$ in a state $\ket{\psi}$
is clearly non-negative, $\bra{\psi}\hat{S}\ket{\psi}\geq0$; equivalently,
its eigenvalues are non-negative throughout. However, $\hat{S}$ having
a \emph{zero} eigenvalue would lead to a contradiction: assume that
there is a normalizable state $\ket{\psi_{0}}$ which $\hat{S}$ annihilates,
$\hat{S}\ket{\psi_{0}}=0$, and expand the associated projector $\hat{S}_{0}=\ket{\psi_{0}}\bra{\psi_{0}}$
in terms of the basis $\hat{Q}^{n}$. The sum of the non-negative
expansion coefficients \begin{equation}
S_{n}=\mbox{Tr}[\hat{S}_{0}\hat{Q}_{n}]=\left|\bra{\psi_{0}}\mathbf{n}_{n}\rangle\right|^{2}\label{eq: explicit S op}\end{equation}
 would vanish since one has \begin{equation}
\sum_{n=1}^{d^{2}}\alpha_{n}S_{n}=\bra{\psi_{0}}\sum_{n=1}^{d^{2}}\alpha_{n}\hat{Q}_{n}\ket{\psi_{0}}=\bra{\psi_{0}}\left(\hat{S}\ket{\psi_{0}}\right)=0\,,\label{eq: vanishing S sum}\end{equation}
which is only possible if each term $S_{n}$ of the sum vanishes individually.
Hence, $\hat{S}_{0}$ must be zero, contradicting the assumption that
$\ket{\psi_{0}}$ is normalizable state. This leaves us with $\hat{S}>0$,
and the operator $\hat{S}$ thus has a unique square root and an inverse,
which is sufficient to complete the CFS-construction. Explicitly,
the resulting family of MIC-POVMs is given by \begin{equation}
\{\hat{E}_{n}=\alpha_{n}\hat{S}^{-\frac{1}{2}}\hat{Q}_{n}\hat{S}^{-\frac{1}{2}}\,,\alpha_{n}>0\,,n=1\ldots d^{2}\}\,.\label{eq: RBSC EVR}\end{equation}
As no analytic expressions for the square roots are available beyond
$d=4$, the POVM just constructed will in general not be in closed
form.

\subsection{MIC-POVMs from the EVR: first case\label{sub: first case}}

As they stand, the expansions given in Eqs. (\ref{eq: identity in basis})
and (\ref{eq: dual evr expansion}) do not define POVMs since neither
the expansion coefficients $\mathbb{I}^{n}$ nor the elements of the
basis $\hat{Q}^{n}$ are generally non-negative. It will be shown
now that minor modifications are sufficient in order to obtain MIC-POVMs. 

Rearrange the terms in (\ref{eq: identity in basis}) in such a way
that a first sum contains expressions with non-negative coefficients
only, $\mathbb{I}^{n_{+}}\geq0$, while a second sum combines the
terms with $\mathbb{I}^{n_{-}}<0$, \begin{equation}
\hat{\mathbb{I}}=\frac{1}{d}\:\sum_{n_{+}=1}^{N_{+}}\mathbb{I}^{n_{+}}\hat{Q}_{n_{+}}-\frac{1}{d}\,\sum_{n_{-}=1}^{N_{-}}\left|\mathbb{I}^{n_{-}}\right|\hat{Q}_{n_{-}},\quad N^{+}+N^{-}=d^{2}\,.\label{eq: sort pos neg}\end{equation}
 Add a $(C/d)$-fold multiple of the identity on both sides, with
$C=\sum_{n_{-}}|\mathbb{I}^{n_{-}}|>0$, to find\begin{equation}
\left(1+\frac{C}{d}\right)\,\hat{\mathbb{I}}=\frac{1}{d}\,\sum_{n_{+}=1}^{N_{+}}\mathbb{I}^{n_{+}}\hat{Q}_{n_{+}}+\frac{1}{d}\,\sum_{n_{-}=1}^{N_{-}}\left|\mathbb{I}^{n_{-}}\right|(\hat{\mathbb{I}}-\hat{Q}_{n_{-}})\,.\label{eq: sort pos neg plus}\end{equation}
This can be written as \begin{equation}
\hat{\mathbb{I}}=\sum_{n_{+}=1}^{N_{+}}\hat{E}_{n_{+}}+\sum_{n_{-}=1}^{N_{-}}\hat{E}_{n_{-}}\equiv\sum_{n=1}^{d^{2}}\hat{E}_{n}\,,\label{eq: new POVM identity}\end{equation}
with $d^{2}$ positive semi-definite operators, $N_{+}$ of which
have rank one and $N_{-}$ have rank $(d-1)$,\begin{equation}
\hat{E}_{n_{+}}=\frac{\mathbb{I}^{n_{+}}}{d+C}\,\hat{Q}_{n_{+}}\geq0\,,\quad\hat{E}_{n_{-}}=\frac{|\mathbb{I}^{n_{-}}|}{d+C}\,(\hat{\mathbb{I}}-\hat{Q}_{n_{+}})\geq0\,.\label{eq: new POVM elements}\end{equation}
 Due to (\ref{eq: new POVM identity}), the operators $\hat{E}_{n_{\pm}}$ form a MIC-POVM, having
a simple physical interpretation: this POVM has $d^{2}$ possible
outcomes, $N_{+}$ of which correspond to finding the system in one
of the coherent states $\ket{\mathbf{n}_{n_{+}}}$, while the remaining
$N_{-}$ outcomes indicate that it is in a state with non-zero component
in a $(d-1)$-dimensional subspace orthogonal to one of the states
$\ket{\mathbf{n}_{n_{-}}}$. The case of a qubit is special since the operators $\hat{E}_{n_{\pm}}$ are of
rank one throughout. The MIC-POVM in (\ref{eq: new POVM identity}) is given
in closed form for any dimension $d$.

\subsection{MIC-POVMs from the EVR: second case\label{sub:second case}}

Not surprisingly, similar modifications enable one to construct a
MIC-POVM out of the elements $\hat{Q}^{n}$ of the dual basis. Write
the expansion (\ref{eq: dual evr expansion}) of the identity as \begin{equation}
\left(1+\frac{\tilde{C}}{d}\right)\,\hat{\mathbb{I}}=\frac{1}{d}\,\sum_{n_{+}=1}^{\tilde{N}_{+}}\hat{Q}^{n_{+}}+\frac{1}{d}\,\sum_{n_{-}=1}^{\tilde{N}_{-}}\left(\hat{Q}^{n_{-}}+|q^{n_{-}}|\hat{\mathbb{I}}\right)\:,\label{eq: identity in dual basis pos neg}\end{equation}
where the first sum contains positive semi-definite operators only,
and the second one takes care of the indefinite ones; the number $q^{n_{-}}<0$
denotes the smallest eigenvalue of $\hat{Q}^{n_{-}}$ and $\tilde{C}$
is the sum of their moduli,\begin{equation}
\tilde{C}=\sum_{n_{-}=1}^{\tilde{N}_{-}}|q^{n_{-}}|>0\,.\label{eq:dual C}\end{equation}
Then, the operators\begin{equation}
\hat{\varepsilon}^{n_{+}}=\frac{1}{d+\tilde{C}}\,\hat{Q}^{n_{+}}\,,\quad\hat{\varepsilon}^{n_{-}}=\frac{1}{d+\tilde{C}}\,(\hat{Q}^{n_{-}}+|q^{n_{-}}|\hat{\mathbb{I}})\,,\label{eq: new dual POVM elements}\end{equation}
are positive semi-definite by construction and give rise to a POVM,
\begin{equation}
\hat{\mathbb{I}}=\sum_{n_{+}=1}^{\tilde{N}_{+}}\hat{\varepsilon}^{n_{+}}+\sum_{n_{-}=1}^{\tilde{N}-}\hat{\varepsilon}^{n_{-}}\equiv\sum_{n=1}^{d^{2}}\hat{\varepsilon}^{n}\,.\label{eq: new dual POVM identity}\end{equation}
As Eq. (\ref{eq: new dual POVM elements}) involves the smallest eigenvalues
of some operators, the resulting POVM is not given in closed form.
It is not difficult to see that this MIC-POVMs is not dual to the
one constructed in the previous section: taking the scalar products
within each basis one has \begin{equation}
\mbox{Tr}\left[\hat{E}_{n}\hat{E}_{n^{\prime}}\right]\geq0\,,\quad\mbox{and}\quad\mbox{Tr}[\hat{\varepsilon}^{n}\hat{\varepsilon}^{n^{\prime}}]\geq0\,,\quad n,n^{\prime}=1\ldots N\,.\label{eq: no duality}\end{equation}
Thus, both sets of operators define their own Gram matrices with only
non-negative entries only; not being diagonal, these matrices cannot
be inverse to each other. The MIC-POVMs in (\ref{eq: new POVM identity})
and (\ref{eq: new dual POVM identity}) are intrinsically different.

\subsection{General MIC-POVMs}

Having gained some experience with the construction of MIC-POVMs,
it becomes obvious how to generalize the CFS-approach. Effectively,
one can both relax the condition of having $d^{2}$ \emph{non-negative}
operators and avoid the appearance of the analytically inaccessible
square root of an operator. Explicitly, it will be shown that \emph{every
set of $d^{2}$ linearly independent hermitean operators acting on
$\mathcal{\mathbb{C}}^{d}$ can be used to define a closed-form MIC-POVM. }

Consider $d^{2}$ hermitean operators $\hat{\kappa}_{n}$ on $\mathcal{\mathbb{C}}^{d}$
with extremal eigenvalues $-\infty<\kappa_{n}^{\pm}<\infty$, not
both of which can be equal to zero simultaneously. Since they satisfy
the inequalities\begin{equation}
\kappa_{n}^{-}\leq\hat{\kappa}_{n}\leq\kappa_{n}^{+}\,,\quad n=1\ldots d^{2}\,,\label{eq: kappa bound}\end{equation}
 the shifted and rescaled operators \begin{equation}
\hat{K}_{n}=\frac{1}{\kappa_{n}^{+}-\kappa_{n}^{-}}\left(\hat{\kappa}_{n}-\kappa_{n}^{-}\hat{\mathbb{I}}\right)\,,\quad n=1\ldots d^{2}\,,\label{eq: scaled kappas}\end{equation}
 are bounded by zero and one, \begin{equation}
0\leq\hat{K}_{n}\leq1\,,\quad n=1\ldots d^{2}\,,\label{eq: K bounds}\end{equation}
as is necessary for the elements of a POVM. The conditions \begin{equation}
\frac{1}{d}\:\mbox{Tr}\left[\hat{K}^{n^{\prime}}\hat{K}_{n}\right]=\delta_{n}^{n^{\prime}}\label{eq: K orthonormality}\end{equation}
 determine a unique dual set of $d^{2}$ operators, $\hat{K}^{n}$.
Hence, there are two expansions of the identity, \begin{equation}
\hat{\mathbb{I}}=\frac{1}{d}\sum_{n=1}^{d^{2}}\mathbb{I}^{n}\hat{K}_{n}=\frac{1}{d}\sum_{n=1}^{d^{2}}\mathbb{I}_{n}\hat{K}^{n}\:,\label{eq: identity in basis general}\end{equation}
where\begin{equation}
\mathbb{I}^{n}=\mbox{Tr}\left[\hat{\mathbb{I}}\hat{K}^{n}\right]\,,\quad\mbox{and}\quad\mathbb{I}_{n}=\mbox{Tr}\left[\hat{\mathbb{I}}\hat{K_{n}}\right]\,.\label{eq: Identity K coeffs}\end{equation}
 As before, some of the coefficients $\mathbb{I}^{n}$ may be negative
and the dual operators $\hat{K}^{n}$ are not necessarily positive
semi-definite. In the first case, follow the procedure described in
Sec. \ref{sub: first case}: effectively replace the operators $\hat{K}_{n_{-}}$
(the ones with \emph{negative coefficients)} by $(\hat{\mathbb{I}}-\hat{K}_{n_{-}})$
leading to set of $d^{2}$ non-negative operators which sum up to
the identity\begin{equation}
\hat{\mathbb{I}}=\sum_{n_{+}=1}^{N_{+}}\hat{\epsilon}_{n_{+}}+\sum_{n_{-}=1}^{N_{-}}\hat{\epsilon}_{n_{-}}\equiv\sum_{n=1}^{d^{2}}\hat{\epsilon}_{n}\,,\label{eq: new general POVM identity}\end{equation}
where 

\begin{equation}
\hat{\epsilon}_{n_{+}}=\frac{\mathbb{I}^{n_{+}}}{d+C}\,\hat{K}_{n_{+}}\,,\quad\hat{\epsilon}_{n_{-}}=\frac{|\mathbb{I}^{n_{-}}|}{d+C}\,(\hat{\mathbb{I}}-\hat{K}_{n_{-}})\,,\label{eq: new general POVM elements}\end{equation}
$C$ being defined as the sum of the moduli of the negative coefficients. 

In the second case, the indefinite dual operators can be made positive
semi-definite by adding an appropriate multiple of the identity to
(\ref{eq: identity in basis general}), in complete analogy to the
procedure presented in Sec. \ref{sub:second case}. As a result, it
has been shown that at least two different MIC-POVMs can be introduced
given any set of $d^{2}$ linearly independent hermitean operators. 

For $d>4$, no analytic expressions for the extremal eigenvalues $\kappa_{n}^{\pm}$
of the operators $\hat{\kappa}_{n}$ exist in general. If a MIC-POVM
in closed form is required, one can resort to the weaker inequalities\begin{equation}
-\|\hat{\kappa}_{n}\|\leq\hat{\kappa}_{n}\leq\|\hat{\kappa}_{n}\|\,,\quad n=1\ldots d^{2}\,,\label{eq: kappa bound 2}\end{equation}
 with any matrix norm $\|\hat{M}\|$. The advantage is that $\|\hat{\kappa}_{n}\|$
can be calculated explicitly once the matrix elements of $\hat{\kappa}_{n}$
are known in some basis. Subsequently, one obtains a modified version
of Eq. (\ref{eq: scaled kappas}),\begin{equation}
\hat{K}_{n}=\frac{1}{2\|\hat{\kappa_{n}}\|}\left(\hat{\kappa}_{n}+\|\hat{\kappa_{n}}\|\hat{\mathbb{I}}\right)\,,\quad n=1\ldots d^{2}\,,\label{eq: new scaled kappas}\end{equation}
ending up with a MIC-POVM given \emph{in closed form}, for any set
of $d^{2}$ linearly independent operators $\hat{\kappa}_{n}$.

\section{Examples: Mic-Povms for a qubit\label{sub: qubit MICs}}

\subsection{Tetrahedral MIC-POVM\label{sub:Tetrahedral-MIC-POVM}}

Consider the operators $\hat{Q}_{n}=\ketbra{\mathbf{n}_{n}}{\mathbf{n}_{n}},n=1\ldots4,$
defined by the following four vectors,\begin{equation}
\mathbf{n}_{1}=(0,0,1)\,,\;\mathbf{n}_{2}=\frac{1}{3}\left(2\sqrt{2},0,-1\right)\,,\;\mathbf{n}_{3}=\frac{1}{3}\left(-\sqrt{2},\sqrt{6},-1\right)\,,\;\mathbf{n}_{4}=\frac{1}{3}\left(-\sqrt{2},-\sqrt{6},-1\right)\,,\label{eq: tetra unit vcs}\end{equation}
 which point to the vertices of a tetrahedron. The projectors $\hat{Q}_{n}$
are linear independent since their Gram matrix is invertible, \begin{equation}
\mathfrak{G}=\frac{1}{3}\left(\begin{array}{cccc}
3 & 1 & 1 & 1\\
1 & 3 & 1 & 1\\
1 & 1 & 3 & 1\\
1 & 1 & 1 & 3\end{array}\right)\,,\quad\mathfrak{G^{-1}}=\frac{1}{4}\left(\begin{array}{cccc}
5 & -1 & -1 & -1\\
-1 & 5 & -1 & -1\\
-1 & -1 & 5 & -1\\
-1 & -1 & -1 & 5\end{array}\right)\,.\label{eq: tetra gram}\end{equation}
The dual operators are given by $\hat{Q}^{n}=(1/2)\sum_{m}\mbox{Tr}[\hat{Q}^{n}\hat{Q}^{m}]\hat{Q}_{m}$,
which gives \begin{equation}
\hat{Q}^{1}=\frac{1}{2}\left(5\hat{Q}_{1}-\hat{Q}_{2}-\hat{Q}_{3}-\hat{Q}_{4}\right)\,,\quad\mbox{etc.}\,,\label{eq: first dual op}\end{equation}
recalling that $\mbox{Tr}[\hat{Q}^{n}\hat{Q}^{m}]\equiv d^{2}\mathfrak{G}^{nm}$,
where $\mathfrak{G}^{nm}=[\mathfrak{G}^{-1}]_{nm}$. This leads to 

\begin{equation}
\hat{\mathbb{I}}=\frac{1}{2}\sum_{n=1}^{4}\hat{Q}_{n}\,,\label{eq: tetra identity}\end{equation}
where $\mathbb{I}^{n}=\mbox{Tr}[\hat{Q}^{n}]=(1/2)(5-1-1-1)=1,n=1\ldots4$,
has been used. Note that the discrete \emph{P}- and \emph{Q}-symbol
of the identity coincide and are both positive, $\mathbb{I}_{n}=\mathbb{I}^{n}=1$.
The resulting operators $\hat{E}_{n}=\hat{Q}_{n}/2\equiv(1/4)(\hat{\mathfrak{\mathbb{I}}}+\mathbf{n}_{n}\cdot\hat{\boldsymbol{\sigma}})$
are exactly those given in Eq. (\ref{eq: tetrahedral POVM}), thus
constituting a MIC-POVM and even a SIC-POVM \cite{flammia+04}. 

The dual operators $\hat{Q}^{n}$ are not positive semi-definite:
consider the expectation value of $\hat{Q}_{1}$ in Eq. (\ref{eq: first dual op})
in the state $\ket{-\mathbf{n}_{1}}$, for example, \begin{equation}
\bra{-\mathbf{n}_{1}}\hat{Q}^{1}\ket{-\mathbf{n}_{1}}=\frac{1}{4}(-3+\mathbf{n}_{1}\cdot\mathbf{n}_{2}+\mathbf{n}_{1}\cdot\mathbf{n}_{3}+\mathbf{n}_{1}\cdot\mathbf{n}_{4})<0\,,\label{eq: negative Qone}\end{equation}
where $|\bra{\mathbf{n}}\mathbf{n}^{\prime}\rangle|^{2}=(1/2)(1+\mathbf{n}\cdot\mathbf{n}^{\prime})$
has been used. Consequently, one would need to apply the procedure
described in Sec. \ref{sub:second case} to determine a second MIC-POVM.

\subsection{Generic MIC-POVMs\label{sub:Generic-MIC-POVMs}}

This section gives an example of the expectation-value representation
for a qubit where the expansion coefficients of the identity with
respect to the original basis are not all positive. Let us consider
three pairwise orthogonal unit vectors $\mathbf{n}_{i}\,,i=1,2,3,$
in $\mathbb{R}^{3}$, and \begin{equation}
\mathbf{n}_{4}=\frac{1}{\sqrt{3}}\left(\mathbf{n}_{1}+\mathbf{n}_{2}+\mathbf{n}_{3}\right)\,.\label{eq: four vectors}\end{equation}
These four vectors clearly do not sum up to zero as is required in
(\ref{eq: tetrahedral POVM}), hence the four projection operators
$\hat{Q}_{n}=\ket{\mathbf{n}_{n}}\bra{\mathbf{n}_{n}}\,,n=1\ldots4\,,$
do not form a POVM. They are, however, linear independent since the
vectors $\mathbf{n}_{n}$ do not lie on a cone which, according to
\cite{heiss+00} is sufficient for a non-singular Gram matrix. 

The procedure outlined in Sec. \ref{sub:second case} associates with
them a unique MIC-POVM, \begin{eqnarray}
\hat{E}_{i} & = & \frac{2}{\sqrt{3}(\sqrt{3}+1)}\ket{\mathbf{n}_{i}}\bra{\mathbf{n}_{i}}\,,\quad i=1,2,3\,,\label{eq: generic POVM}\\
\hat{E}_{4} & = & \frac{2}{(\sqrt{3}+1)}\ket{-\mathbf{n}_{4}}\bra{-\mathbf{n}_{4}}\,.\label{eq: generic POVM 2}\end{eqnarray}
 The derivation of this result is simplified by the the fact \cite{heiss+00}
that one can express the expansion coefficients of the identity in
the form \begin{equation}
\mathbb{I}^{n}=\frac{4}{1+\mathbf{f}^{n}\cdot\mathbf{n}_{n}}\:,\quad n=1\ldots4\,,\label{eq: f coeffs}\end{equation}
where the vector $\mathbf{f}^{1}\in\mathbb{R}^{3}$ is determined
by \begin{equation}
\mathbf{f}^{1}=-\frac{\mathbf{n}_{2}\wedge\mathbf{n}_{3}+\mathbf{n}_{3}\wedge\mathbf{n}_{4}+\mathbf{n}_{4}\wedge\mathbf{n}_{2}}{(\mathbf{n}_{2}\wedge\mathbf{n}_{3})\cdot\mathbf{n}_{4}}\,,\label{eq: f vector}\end{equation}
 and the other three vectors follow from this relation by cyclic permutation
of the indices 1 through 4. A straightforward but still lengthy calculation
leads to \begin{eqnarray}
\mathbb{I}^{i} & = & \frac{4}{\sqrt{3}(\sqrt{3}-1)}>0\,,\quad i=1,2,3\,,\label{eq: generic I i}\\
\mathbb{I}^{4} & = & \frac{4}{(1-\sqrt{3})}<0\,,\label{eq: generic I4}\end{eqnarray}
i.e. there is one negative coefficient which needs to be eliminated by
adding a multiple of the identity to the expansion of the identity.
Apply now the method outlined in Sec. \ref{sub: first case}, and
you will find the MIC-POVM specified in (\ref{eq: generic POVM},\ref{eq: generic POVM 2}).
A second MIC-POVM could be obtained from the dual basis but no further
insight is to be gained form its explicit form.

\section{Summary and Conclusions\label{sec:Conclusions}}

Starting from the expectation-value representation of quantum mechanics
in $d$-dimensional Hilbert spaces, new simple POVMs with $d^{2}$
elements have been introduced which are informationally complete. Mathematically
speaking, the elements of these POVMs provide a basis in the Hilbert-Schmidt
space of operators acting on $\mathcal{\mathbb{C}}^{d}$ while, from
a physical point of view, they are suited to reconstruct unknown quantum
states if an arbitrarily large number of systems in the same state
are available. Repeated measurements with such a POVM generate $d^{2}$
probabilities which are in a one-to-one correspondence with a density matrix
$\hat{\rho}$. Since any set of $d^{2}$ linearly independent operators
can be used as a starting point, a wide range of possibilities opens
up to construct MIC-POVMs most suited for the application at hand. 

It seems worthwhile to finally point out how to explicitly write down
a density matrix $\hat{\rho}$ once the probabilities \begin{equation}
p_{n}(\hat{\rho})=\mbox{Tr}\left[\hat{\rho}\hat{E}_{n}\right]\in[0,1]\,,\quad n=1\ldots d^{2}\,,\label{eq: measured probs}\end{equation}
associated with a MIC-POVM $\hat{E}_{n},n=1\ldots d^{2}$, have been
measured. The most direct approach invokes the dual operators $\hat{E}^{n},$
defined by the equivalent of the condition (\ref{eq:duality}). Once
these operators have been found, the density matrix is given by

\begin{equation}
\hat{\rho}=\frac{1}{d}\sum_{n=1}^{d^{2}}p_{n}(\hat{\rho})\hat{E^{n}}\,.\label{eq: general reconstruction}\end{equation}
Formally, this result is very similar to Eq. (\ref{eq: EVR expansion}), its equivalent in the expectation-value
representation. However,
there is a fundamental difference since the numbers $p_{n}(\hat{\rho})$ are
`honest' probabilities emerging from an experiment performed with
a \emph{single} apparatus while the probabilities required for the
expectation-value representation are obtained from running $d^{2}$
different experiments.

\end{document}